\input harvmac

\input epsf

\newcount\figno
\figno=0 
\def\fig#1#2#3{
\par\begingroup\parindent=0pt\leftskip=1cm\rightskip=1cm\parindent=0pt
\baselineskip=11pt
\global\advance\figno by 1
\midinsert
\epsfxsize=#3
\centerline{\epsfbox{#2}}
\vskip 12pt
{\bf Fig.\ \the\figno: } #1\par
\endinsert\endgroup\par
}
\def\figlabel#1{\xdef#1{\the\figno}}
\def\encadremath#1{\vbox{\hrule\hbox{\vrule\kern8pt\vbox{\kern8pt
\hbox{$\displaystyle #1$}\kern8pt}
\kern8pt\vrule}\hrule}}
\def\adag{a^{\dagger}}

\font\cmss=cmss10
\font\cmsss=cmss10 at 7pt
\def\rlx{\relax\leavevmode}
\def\inbar{\vrule height1.5ex width.4pt depth0pt}
\def\IN{\relax{\rm I\kern-.18em N}}
\def\IP{\relax{\rm I\kern-.18em P}}
\def\ZZ{\rlx\leavevmode\ifmmode\mathchoice{\hbox{\cmss Z\kern-.4em Z}}
 {\hbox{\cmss Z\kern-.4em Z}}{\lower.9pt\hbox{\cmsss Z\kern-.36em Z}}
 {\lower1.2pt\hbox{\cmsss Z\kern-.36em Z}}\else{\cmss Z\kern-.4em
 Z}\fi}
\def\IZ{\relax\ifmmode\mathchoice
{\hbox{\cmss Z\kern-.4em Z}}{\hbox{\cmss Z\kern-.4em Z}}
{\lower.9pt\hbox{\cmsss Z\kern-.4em Z}}
{\lower1.2pt\hbox{\cmsss Z\kern-.4em Z}}\else{\cmss Z\kern-.4em
Z}\fi}
\def\IZ{\relax\ifmmode\mathchoice
{\hbox{\cmss Z\kern-.4em Z}}{\hbox{\cmss Z\kern-.4em Z}}
{\lower.9pt\hbox{\cmsss Z\kern-.4em Z}}
{\lower1.2pt\hbox{\cmsss Z\kern-.4em Z}}\else{\cmss Z\kern-.4em
Z}\fi}

\def\narrowplus{\kern -.04truein + \kern -.03truein}
\def\narrowminus{- \kern -.04truein}
\def\narrowminussub{\kern -.02truein - \kern -.01truein}

\def\Sym{{\rm Sym}}
\def\Hilb{{\rm Hilb}}
\def\zb{{\bar z}}
\def\wb{{\bar w}}
\def\Gr{{\rm Gr}}
\def\phh{{\hat\phi}}
\def\ph{{\phi}}
\def\t{{\theta}}
\def\l{{\lambda}}
\def\w{{\omega}}
\def\s{{\sigma}}
\def\frac#1#2{{#1\over #2}}

\def\CE{{\cal E}}
\def\CM{{\cal M}}
\def\IZ{\relax\ifmmode\mathchoice
{\hbox{\cmss Z\kern-.4em Z}}{\hbox{\cmss Z\kern-.4em Z}}
{\lower.9pt\hbox{\cmsss Z\kern-.4em Z}}
{\lower1.2pt\hbox{\cmsss Z\kern-.4em Z}}\else{\cmss Z\kern-.4em
Z}\fi}
\def\IC{{\relax\,\hbox{$\inbar\kern-.3em{\rm C}$}}}
\def\p{\partial}
\font\cmss=cmss10 \font\cmsss=cmss10 at 7pt
\def\IR{\relax{\rm I\kern-.18em R}}
\def\ra{\rangle}
\def\la{\langle}

\nref\wsft{E. Witten,
``Noncommutative Geometry And String Field Theory,''
Nucl.\ Phys.\  {\bf B268} (1986) 253.}

\nref\sw{N. Seiberg and E. Witten,
``String Theory and Noncommutative Geometry,''
JHEP {\bf 09} (1999) 032, [hep-th/9908142].}

\nref\wittach{E. Witten,
``Noncommutative Tachyons and String Field Theory,''
[hep-th/0006071].}

\nref\schn{M. Schnabl,
``String field theory at large B-field and noncommutative geometry,''
JHEP {\bf 11} (2000) 031,
[hep-th/0010034].}

\nref\gms{R. Gopakumar, S. Minwalla and A. Strominger,
``Noncommutative solitons,'' JHEP {\bf 05} (2000) 020,
[hep-th/0003160].}

\nref\rms{K. Dasgupta, S. Mukhi and G. Rajesh,
``Noncommutative tachyons,'' JHEP {\bf 06} (2000) 022,
[hep-th/0005006].}

\nref\hklm{J. A. Harvey, P. Kraus, F. Larsen and E. J. Martinec,
``D-branes and Strings as Non-commutative Solitons,''
JHEP {\bf 07} (2000) 042,
[hep-th/0005031].}

\nref\senb{A. Sen,
``Descent Relations Among Bosonic D-branes,''
 Int.\ J. Mod.\ Phys.\ { \bf A14} (1999) 4061,
[hep-th/9902105].}

\nref\senc{A. Sen,
 ``Universality of the Tachyon Potential,''
JHEP {\bf 12} (1999) 027,
[hep-th/9911116].}

\nref\poly{A. P. Polychronakos,
``Flux tube solutions in noncommutative gauge theories,''
Phys.\ Lett.\ {\bf B495} (2000) 407,
[hep-th/0007043].}

\nref\bak{D. Bak,
``Exact Solutions of Multi-Vortices and False Vacuum Bubbles in
Noncommutative Abelian-Higgs Theories,''
Phys.\ Lett.\ {\bf B495} (2000) 251,
[hep-th/0008204].}

\nref\agms{M. Aganagic, R. Gopakumar, S. Minwalla and A. Strominger,
``Unstable solitons in noncommutative gauge theory,''
JHEP {\bf 04} (2001) 001,
[hep-th/0009142].}

\nref\gn{D. J. Gross and N. A. Nekrasov,
``Solitons in noncommutative gauge theory,''
{\bf 03} (2001) 044,
[hep-th/0010090].}

\nref\jmw{D. P. Jatkar, G. Mandal and S. R. Wadia,
``Nielsen-Olesen Vortices in Noncommutative Abelian Higgs Model,''
JHEP {\bf 09} (2000) 018,
[hep-th/0007078].}

\nref\baklee{D. Bak and K. Lee,
``Elongation of Moving Noncommutative Solitons,''
Phys.\ Lett.\ {\bf B495} (2000) 231,
[hep-th/0007107].}

\nref\mak{A. S. Gorsky, Y. M. Makeenko and K. G. Selivanov,
``On noncommutative vacua and noncommutative solitons,''
Phys.\ Lett.\ {\bf B492} (2000) 344,
[hep-th/0007247].}

\nref\zhou{C. Zhou,
``Noncommutative scalar solitons at finite $\t$,''
[hep-th/0007255].}

\nref\solov{A. Solovyov,
``On Noncommutative Solitons,''
Mod.\ Phys.\ Lett.\ {\bf A15} (2000) 2205,
[hep-th/0008199].}

\nref\hkl{J. A. Harvey, P. Kraus and F. Larsen,
``Exact noncommutative solitons,''
JHEP {\bf 12} (2000) 024,
[hep-th/0010060].}

\nref\dur{B. Durhuus, T. Jonsson and R. Nest,
``Noncommutative scalar solitons: existence and nonexistence,''
Phys.\ Lett.\ {\bf B500} (2001) 320,
[hep-th/0011139].}

\nref\miaoli{M. Li,
``Quantum Corrections to Noncommutative Solitons,''
[hep-th/0011170].}

\nref\kiem{Y. Kiem, C. Kim and Y. Kim,
``Noncommutative Q-balls,''
[hep-th/0102160].}

\nref\jackson{M. G. Jackson, ``The Stability of
Noncommutative Scalar Solitons,''
[hep-th/0103217].}

\nref\lru{U. Lindstr\"om, M. Ro\v{c}ek and R. von Unge,
``Non-commutative Soliton Scattering,''
JHEP {\bf 12} (2000) 004,
[hep-th/0008108].}

\nref\manton{N. S. Manton, ``A remark on the scattering of BPS
monopoles,'' Phys.\ Lett.\ {\bf B110} (1982) 54.}

\nref\bars{I. Bars, H. Kajiura, Y. Matsuo and T. Takayanagi,
``Tachyon Condensation on Noncommutative Torus,''
Phys.\ Rev.\ {\bf D63} (2001) 086001,
[hep-th/0010101].}

\nref\ss{E. M. Sahraoui and E. H. Saidi,
``Solitons on compact and noncompact spaces in large noncommutativity,''
[hep-th/0012259].}

\nref\mm{E. J. Martinec and G. Moore,
``Noncommutative Solitons on Orbifolds,''
[hep-th/0101199].}

\nref\blp{D. Bak, K. Lee and J.-H. Park,
``Noncommutative vortex solitons,''
[hep-th/0011099].}

\nref\lty{K. Lee, D. Tong and S. Yi,
``The moduli space of two U(1) instantons on noncommutative
$R^4$ and $R^3\times S^1$,''
[hep-th/0008092].}

\nref\ly{K. Lee and P. Yi,
``Quantum Spectrum of Instanton Solitons 
in five dimensional noncommutative U($N$) theories,''
[hep-th/9911186].}

\nref\perelomov{A. M. Perelomov,
``On the completeness of a system of coherent states,''
Teor.\ Mat.\ Fiz. {\bf 6} (1971) 213.
}

\nref\bbgk{
V. Bargmann, P. Butera, L. Girardello and J. R. Klauder,
``On the completeness of the coherent states,''
Rep. on Math. Phys. {\bf 2} (1971) 221.
}

\nref\bgz{
H. Bacry, A. Grossman and J. Zak,
``Proof of completeness of lattice states in the $kq$ representation,''
Phys. Rev. {\bf B12} (1975) 1118.
}

\nref\nakajima{H. Nakajima, ``Lectures on Hilbert schemes of points
on surfaces,'' American Mathematical Society University Lecture
Series, 1999.}

\nref\furuuchi{K. Furuuchi, ``Instantons on Noncommutative
$R^4$ and Projection Operators,'' Prog.\ Theor.\ Phys.\ {\bf 103}
(2000) 1043, [hep-th/9912047].}

\nref\nekrasov{N. A. Nekrasov, ``Noncommutative instantons revisited,''
[hep-th/0010017].}

\nref\bn{H. W. Braden, N. A. Nekrasov, ``Instantons, Hilbert Schemes
and Integrability,'' [hep-th/0103204].}

\lref\hlru{L. Hadasz, U. Lindstr\"om, M. Ro\v{c}ek, R. von Unge,
``Noncommutative Multisolitons: Moduli Spaces, Quantization, Finite $\t$
Effects, and Stability,''
[hep-th/0104017].}

\lref\lecht{O. Lechtenfeld, A. D. Popov, B. Spendig,
``Noncommutative solitons in open N=2 string theory,''
[hep-th/0103196].}

\lref\zak{J. Zak,
in {\sl Solid State Physics}, edited by H. Ehrenreich, F. Seitz, and
D. Turnbull (Academic, New York, 1972), Vol. 27.
}

\lref\vn{
J. von Neumann, {\sl Mathematical Foundations of Quantum Mechanics},
Princeton (1955).}

\lref\str{Talk by R. Gopakumar at Strings 2001, Mumbai,
\vskip 0.0001in
{\tt http://theory.tifr.res.in/strings/Proceedings/gkumar/}.}

\Title{\vbox{\baselineskip12pt\hbox{hep-th/0103256}\hbox{}
\hbox{}}}{On Noncommutative Multi-solitons}

\centerline{Rajesh Gopakumar, Matthew Headrick
and Marcus Spradlin}
\bigskip\centerline{Jefferson Physical Laboratory}
\centerline{Harvard University}
\centerline{Cambridge MA 02138}
\centerline{E-mail: {\tt headrick@physics.harvard.edu}}
\smallskip

\vskip .3in \centerline{\bf Abstract}

\noindent
We find the moduli space of multi-solitons in noncommutative scalar
field theories at large $\t$, in arbitrary dimension. The existence of a
non-trivial moduli space at leading order in $1/\t$ is a consequence of a
Bogomolnyi bound obeyed by the kinetic energy of the $\t=\infty$ solitons.
In two spatial dimensions, the parameter space for $k$ solitons is a
K\"ahler de-singularization of the symmetric product $(\IR^2)^k/S_k$.  We
exploit the existence of this moduli space to construct solitons on
quotient spaces of the plane: $\IR^2/\IZ_k$, cylinder, and $T^2$.
However, we show that tori of area less than or equal to $2\pi\t$ do not
admit stable solitons.  In four dimensions the moduli space provides an
explicit K\"ahler resolution of $(\IR^4)^k/S_k$. In general spatial
dimension $2d$, we show it is isomorphic to the Hilbert scheme of $k$
points in $\IC^d$, which for $d>2$ (and $k>3$) is not smooth and can
have multiple branches.

\Date{March 2001}

\listtoc
\writetoc

\newsec{Introduction}

Among the many reasons to study noncommutative field theories is that they
might be useful for studying stringy behavior in a controlled
manner. In the context of string theory, the noncommutativity is a remnant
of the noncommutative geometry of open string field theory 
\refs{\wsft-\schn}.
An interesting class of excitations in the field theory 
are noncommutative solitons \gms. 
The existence and form of these classical
solutions are fairly independent of the details of the theory, making
them useful probes of stringy behavior.
Coincident solitons
exhibit a non-abelian enhancement of the zero mode spectrum, and
in fact these solitons are the D-branes of string theory
manifested in a field theory limit
while still capturing many of their stringy features. 
Note that, in contrast, D-branes do
not appear as finite energy excitations in conventional (commutative)
field theory limits of string theory.

The fact that one can see D-branes in a simpler field theoretic context
has had important consequences, primarily in the study of tachyon
condensation in open string field theory \refs{\rms,\hklm}. 
Many of Sen's conjectures \refs{\senb,\senc} on this subject
have been beautifully confirmed using properties of noncommutative
solitons. 
Similarly, unstable solitons in noncommutative gauge
theory \refs{\poly,\bak} can be interpreted as codimension two (or higher) 
D-branes localized on other D-branes \refs{\agms,\gn}. 
The gauge theory successfully reproduces the dynamics of 
tachyon condensation in this system. Other studies of scalar solitons 
include \refs{\jmw-\jackson}.

In this paper we will begin to investigate some aspects of how D-branes see
spacetime by exploring the moduli space of solitons in noncommutative scalar
field theory in $2d+1$ dimension. 
The study of multi-solitons itself turns out to
have many interesting features. At $\t=\infty$, because of the presence
of the U$(\infty)$ symmetry in the dominant potential energy term, 
there is an infinite dimensional moduli space of
solutions corresponding to arbitrary hermitian projection operators on
a Hilbert space $\CH$.
Since this symmetry is not preserved by the kinetic term, one
might expect the degeneracy to be completely lifted when one
includes this leading $1/\t$ correction. Surprisingly, we find
that there remains a smaller yet non-trivial moduli space at this order,
which roughly corresponds to individual gaussian solitons free to roam the
plane. The lack of a force between static solitons at this order in
$1/\t$ is due to a Bogomolnyi bound for the kinetic energy of projection
operators. 

This moduli space, however, is not protected by any symmetry, and is
lifted by a classical effective
potential which is generated at the next order in $1/\theta$, and presumably
by quantum corrections as well. The
classical effective potential (which is bounded both above and below)
leads to an attractive force
among the solitons. Thus in the true (exact)
moduli space all $k$ solitons
are coincident. Nevertheless, the approximate moduli space is a useful
description of the dynamics of multiple solitons within a certain range of 
energies \manton .

Why is this moduli space interesting if it is only an approximate one? 
The answer is twofold. Firstly, this moduli space is very closely related 
to the symmetric product spaces $(\IR^{2d})^k/S_k$, which arise
in the study of supersymmetric vacua of D$p$-D$p'$ systems in string
theory. In fact, we will see that our moduli space (for a scalar 
field theory with noncommutativity in $2d$ spatial directions)
is precisely the same as
the so-called Hilbert scheme of $k$ points in $\IR^{2d}$. 
This is, for $d\leq 2$, a smooth resolution of the symmetric product space
which is potentially 
singular when points coincide. For the case of $d=2$ we see that
the solitons resolve the singularity in an interesting
way: when any two of them are brought together, the final configuration
depends on the complex direction in the relative $\IR^4$ by which 
they approach each other, so that there is a hidden $\IP^1$ at the
putative singular point. This is also exactly what D-branes see 
as a stringy resolution of the singularity.
For $d>2$, interestingly, the
Hilbert scheme is not a manifold in any sense and has exotic branches
which also show up in brane systems. In our context, these branches reflect the
fact that coincident solitons in
higher dimension have a large number of moduli associated with the shape
of the lump solution.
 
The second reason why these approximate moduli spaces are interesting is that
they enable us to
construct infinite arrays of solitons in $\IR^{2d}$ that respect some
discrete symmetry. 
Such an array can be viewed as a single soliton on the
quotient space. 
The lowest energy stable solutions on the quotient space---the analogues of
the gaussian soliton on $\IR^2$---are constructed this way.
For the torus we
encounter the somewhat unexpected fact that solitons do not exist when
the area is less than or equal to $2\pi\theta$. 

The moduli spaces also inherit an interesting geometric structure
from the noncommutativity.
For instance, in the simplest case of 
two spatial dimensions, the $k$ soliton moduli space is---as a
complex manifold---simply $\Sym^k(\IR^2) \equiv (\IR^2)^k/S_k$, the
symmetric product of the single soliton moduli space $\IR^2$.
Geometrically, however, the solitons smooth out the conical singularities
that occur on $\Sym^k(\IR^2)$ where two or more points come together,
as the explicit K\"ahler metric shows.
(The K\"ahler metric on the moduli space
of $k=2$ solitons has appeared in the work
of Lindstr\"om et al. \lru .) Similarly, the smooth Hilbert scheme
resolution of $\Sym^k(\IR^4)$ also inherits a K\"ahler metric, distinct from
the hyperk\"ahler one which arises in the case of instanton physics.

That noncommutative solitons have such a rich structure in their moduli
space is very encouraging and makes them worthy of further exploration. 
The way apparent singularities are resolved is very stringy, and
it is interesting that the noncommutative
algebra of projection operators sees the resolved geometry in a simple
way, which the commutative algebra of functions cannot.
This is perhaps a clue 
that noncommutative algebras might play a fundamental role in
understanding geometry in string theory. 

Some of the results in this paper were announced at Strings 2001 \str.  
The paper \mm\ by E. Martinec and G. Moore, which has appeared in the
meanwhile, has overlap with the discussion in section 5. K. Lee and
collaborators have studied the resolution of moduli spaces of instantons
and vortices on noncommutative spaces in \refs{\blp-\ly}.

\newsec{The moduli space at infinite $\t$}

In this section we briefly review the construction of solitons
in 2+1 dimensional noncommutative scalar field theory at large $\t$.
The moduli space at $\t = \infty$ (defined precisely below) is isomorphic
to the space of projection operators\foot{
Throughout this paper, by ``projection operator'' we mean ``hermitian
projection operator.''} on an infinite-dimensional
Hilbert space. We review the relevant mathematical details about
the geometry of this space (known as the Grassmannian).

\subsec{Solitons and projection operators}

Until the last section of this paper, we work in a 2+1 dimensional scalar
field theory with spacelike noncommutativity:
\eqn\action{
S = \int dt\,d^2\!w\left({1\over2}{\dot\ph}^2-
\p_w\ph\p_\wb\ph-m^2 V(\ph)_\star \right),
}
where $w = (x^1+ix^2)/\sqrt{2}$. The subscript on the potential
indicates that it is evaluated with the Moyal star product, with
noncommutativity parameter $\t^{w\wb} = -i\t$ (for details, see \gms). It
will be convenient to
let the factor $m^2$ multiply the entire potential, so we assume
$V''(0)=1$. The existence of stable solitons in this theory depends on the
potential having a second, local minimum, which we put at
$\ph = \l$, with $V(\l) > V(0) = 0$ (figure 1). 
The energy functional for static configurations, $E[\ph] = \int
d^2w(\p_w\ph\p_\wb\ph+m^2 V(\ph)_\star)$, may be conveniently rewritten
using the Weyl-Moyal correspondence,
$w\to\sqrt{\t}a$, $\p_w \to -{1\over \sqrt\t}[\adag,\cdot\,]$, etc., which
maps the star product into operator multiplication in the 
Hilbert space $\CH$ of a 1-dimensional particle:
\eqn\enfun{
E[\phh,\t m^2] 
= 2\pi\,\Tr_{\CH}\left([a,\phh][\phh,\adag]+\t m^2 V(\phh)\right).
}

\fig{
The potential for $\ph$ is assumed to have a global minimum at $\ph=0$ and a
local minimum at $\ph=\l$.
}{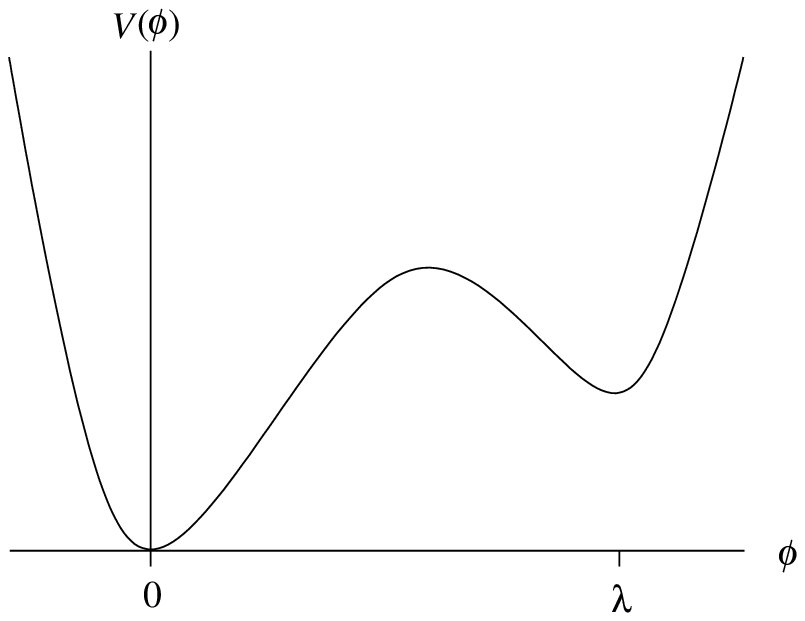}{2.1in}

Fixing the function $V(\ph)$, it is the dimensionless parameter $\t
m^2$ that controls the relative importance of the kinetic and potential
terms in \enfun. Exact solutions to the equation of motion are known in
the limit $\t m^2\to\infty$, when the kinetic term may be neglected
compared to the potential term \gms. In this section we describe the
moduli space of such solitons. In the next section we include the kinetic
term as a perturbation, and see how it lifts the moduli
space down to a much smaller but still non-trivial one.

To define the theory in the limit $\t m^2\to\infty$ we must rescale the
energy:
\eqn\ezero{
E_0[\phh] \equiv \lim_{\t m^2\to\infty} {1\over\t m^2} E[\phh,\t m^2] 
= 2\pi\,\Tr\,V(\phh).
}
Stable solutions to the resulting equation of motion
$V^{\prime}(\phh)=0$ take the form 
\eqn\proj{
\phh = \l P,
}
where $P$ is any projection operator.
The energy of such a solution is $E_0=2\pi kV(\l)$, where $k$ is the rank
of $P$, so we will assume
for now that this rank is finite. (In section 5 we will also discuss
projection operators of infinite rank.)
If one interprets the rank one solutions as single solitons, then 
the rank $k$ solutions may be thought of as corresponding to $k$ 
non-interacting solitons, 
each of energy $2\pi V(\l)$. This interpretation will become more
meaningful in the next section.

Note that \ezero\ has an invariance under arbitrary unitary
transformations $U$ of the Hilbert space, under
which $\phh \to U \phh U^\dagger$.
Since any two projection operators of rank $k$ can be
continuously connected by U($\infty$) transformations, there is an infinite
dimensional moduli space of solutions with energy $2\pi kV(\l)$. 
In fact, the rank $k$ projection operators on $\CH$ (or
equivalently, the $k$-dimensional hyperplanes in $\CH$) form a manifold
known as the Grassmannian $\Gr(k,\CH)$, which can also be described as the
coset space
\eqn\grass{
{{\rm U}(\infty)\over{\rm U}(k)\times{\rm U}(\infty-k)},
}
where U($\infty$) acts on the entire space, while U($\infty-k$) acts only
on the orthogonal complement of a $k$-dimensional hyperplane. 
The U($\infty$) symmetry protects the moduli space against
corrections.

In the next subsection,
we describe the geometry of the Grassmannian, partly because of its
interest as the moduli space of solitons in the limit $\t m^2\to\infty$,
but more importantly because this will give us the tools to
study the geometry of the moduli space of solitons at finite $\t m^2$,
which is a submanifold of the Grassmannian.

\subsec{Geometry of the Grassmannian}

The Grassmannian has a natural complex structure, which it
inherits from $\CH$. Points (vectors) in $\CH$ may be parametrized by a
(infinite) set of holomorphic coordinates $z_a$; these could be the
coefficients of the vector in some particular basis.
If we now have a set of $k$ linearly independent vectors 
$|\psi_i\ra\in\CH$ which depend holomorphically on the
$z_a$, then the hyperplane they span is also considered to depend
holomorphically on the $z_a$.

The Grassmannian also has a natural K\"ahler
structure, which can be computed explicitly as follows.\foot{
For the mathematical cognoscenti: the natural K\"ahler
form may be obtained as the curvature of a certain line bundle.
Let $E$ be the tautological bundle whose fiber over a point
in $\Gr(k,\CH)$ is simply the $k$-dimensional hyperplane that it is.
The inner product $\la \cdot | \cdot \ra$ on $E$ induces
a natural metric on the determinant bundle $\det(E)$: the
norm of a section $\psi = |\psi_1 \ra \wedge \cdots \wedge |\psi_k\ra$ of
$\det(E)$ is
simply $||\psi||^2 = \det(\la \psi_i|\psi_j\ra)$.
The curvature form $i \p_a \p_{\bar{b}} \ln \det(\la \psi_i|\psi_j\ra)$
of this bundle is the natural K\"ahler form on $\Gr(k,\CH)$.
}
Defining the matrix
\eqn\hij{
h_{ij} \equiv \la\psi_i|\psi_j\ra
}
and its inverse $h^{ij}$, the hermitian operator that projects onto the
hyperplane spanned by the $|\psi_i\ra$ is
\eqn\pdef{
P = |\psi_i\ra h^{ij}\la\psi_j|.
}
The matrix $h_{ij}$ is the metric on the image of $P$.
It is straightforward to show that the metric on the Grassmannian is K\"ahler,
\eqn\metric{
g_{a \bar{b}} \equiv \Tr(\p_a P \p_{\bar{b}} P)= \p_{a} \p_{\bar{b}} K,
}
and the K\"ahler potential is given simply by
\eqn\kahpot{
K = \ln\det (h_{ij}).
}
There is an
ambiguity in choosing the $|\psi_i\ra$, with any two choices being
related by a GL($k$) matrix that depends holomorphically on the
coordinates. The respective K\"ahler potentials will
differ by a holomorphic plus an anti-holomorphic function, leaving the
metric unchanged.

Although the derivation of the moduli space metric has been presented as
a mathematical triviality,
we emphasize that
the physical moduli space metric, which arises as the kinetic term
for time-dependent moduli $z_a(t)$, differs from \metric\ only by an overall
factor.  After rescaling the energy as in \ezero, we find the physical metric
\eqn\pmodmet{
{1 \over \t m^2} \int d^2 w\,  \p_a \phi \p_{\bar b}
\phi = {2 \pi \lambda^2 \over m^2} \Tr(\p_a P \p_{\bar{b}} P)
= {2 \pi \lambda^2 \over m^2} \p_a \p_{\bar{b}} K,
}
with $K$ given by the formula \kahpot.

\newsec{The moduli space at finite $\t$}

The U($\infty$) symmetry that protects the moduli space in
the limit $\t m^2 \to \infty$ is broken in the scalar theory
at finite $\t m^2$ by the kinetic term.  One might expect
that the infinite dimensional moduli space $\Gr(k,\CH)$
discussed in the previous section
is completely lifted at finite $\t$.
Remarkably, we find that despite the lack of symmetry,
the leading $1/(\t m^2)$ correction to the
energy satisfies a Bogomolnyi-like bound which preserves 
a finite dimensional subspace of the space of projection
operators of rank $k$.  We show that the remaining moduli space
$\CM_k$ corresponds to $k$ individual solitons which are free to roam
the plane, and that $\CM_k$ has a K\"ahler metric which
is smooth even when the solitons come together.

\subsec{The Bogomolnyi bound}

In a perturbation expansion in $1/(\t m^2)$,
\eqn\perturb{\eqalign{
\phh &= \phh_0 + {1\over \t m^2}\phh_1 + \cdots, \cr
E &= \t m^2 E_0 + E_1 + {1\over \t m^2}E_2 + \cdots,
}}
the first correction to the energy is just the kinetic term:
\eqn\eone{
E_1[\phh_0] =
2\pi\,\Tr[a,\phh_0][\phh_0,\adag].
}
Due to the fact that $V'(\phh_0) = 0$, $E_1$ is independent
of the correction $\phh_1$.
$E_1$ will act like a potential on the space of minima of
$E_0$ described in the previous section. The minima of this potential will
form the moduli space at finite $\t m^2$, which
may then be further corrected at
higher orders in $1/(\t m^2)$.

A reasonable guess would be that a minimum of the kinetic energy is
achieved only by rotationally symmetric solutions.
Rotations (about the origin, for simplicity) are generated by the harmonic
oscillator Hamiltonian $\adag a+1/2$, so rotational symmetry translates,
for operators, into being diagonal in a basis of harmonic oscillator
eigenstates $|n\ra$. Indeed, it was shown in \gms\ that any sum of
operators of the 
form $\l|n\ra\la n|$ extremizes $E_1$ (within the infinite $\t m^2$ moduli
space), while the operator
\eqn\eonemin{
\l\sum_{n=0}^{k-1}|n\ra\la n|
}
was conjectured to minimize it.\foot{
The authors of \gms\ showed that $\l |0\ra \la 0|$
indeed minimizes $E_1$, and that infinitesimal unitary transformations
which mix $|0\ra$ and $|1\ra$ are zero modes of $E_1$ around that extremum,
thus  establishing the conjecture for $k=1,2$.
}
In particular, for $k=1$ this would be achieved by the gaussian
soliton $\l |0\ra \la 0|$.
Using the (exact) translational symmetry of the theory to center such a
solution at any point on the plane, the
moduli space (for any given $k$) appears to be simply the plane itself,
and the moduli space dynamics completely trivial.

However, the story is not so simple. It turns out that there are
non-rotationally symmetric minima of $E_1$.
Indeed, the full moduli space $\CM_k$ has an
interesting structure, large enough to allow non-trivial
dynamics. This unexpected fact is a consequence, not of any symmetry
possessed by $E_1$, but rather of a Bogomolnyi-like bound that it
satisfies:
\eqn\bogom{
E_1[\phh_0] = 2\pi\l^2\,\Tr[a,P][P,\adag]
= 2\pi\l^2\,\Tr\left(P+2F(P)^\dagger F(P)\right)
\ge 2\pi\l^2k
}
where
\eqn\adef{
F(P) \equiv (1-P)aP.
}
The bound is saturated when $F(P)=0$, in other words when 
the image of $P$ is an
invariant subspace of the operator $a$.\foot{
A similar equation has arisen in a different context in \lecht.
}
The projection operators
satisfying this condition define a subspace $\CM_k$ of the Grassmannian
$\Gr(k,\CH)$.  In the next subsection we will see
that it is a finite dimensional subspace, and that the field configurations
corresponding to these projection operators
have a natural interpretation in terms of separated solitons. 
In the following subsection we will then investigate
the geometry of $\CM_k$, showing in particular that it is smooth.

\subsec{Topology}

Starting with the simplest case of $k=1$, it is clear that any
1-dimensional invariant subspace of $a$ must be spanned by an
eigenstate of that operator, i.e. by a coherent
state. We use the non-normalized coherent states $|z\ra \equiv
e^{z\adag}|0\ra$, their virtue for our purposes being
that they depend holomorphically upon the eigenvalue $z$, so that they
form a complex submanifold of $\CH$. This in turn implies that the moduli
space $\CM_1$ is a (1-dimensional) complex submanifold of $\Gr(1,\CH)$.
It is therefore K\"ahler, and in fact the metric \metric\ is simply
$ds^2=dz\,d\zb$.  The solution
\eqn\projone{
\phh_z=
\l{|z\ra\la z|\over\la z|z\ra}
}
maps back under the Weyl-Moyal correspondence to the function
\eqn\solone{
\phi_z(w,\wb) =
2\l e^{-2|w/\sqrt{\t}-z|^2}.
}
This is just a translated gaussian soliton localized around
$w=\sqrt{\t}z$. Its moduli space is naturally isomorphic to the physical
plane.

At higher $k$, the most obvious way to construct an invariant subspace of
$a$ is as a direct sum of such 1-dimensional invariant subspaces, each
spanned by a different coherent state $|z_i\ra$. We can think of this as
$k$ solitons, each with independent collective coordinate $z_i$. (Indeed,
if the $z_i$ are far from each other, then the respective coherent states
are nearly orthogonal and the corresponding field configuration is
approximately the sum of distant Gaussian solitons \solone.)
The moduli space is, at least naively, 
the $k$-fold symmetric product of the single-soliton moduli
space, $\Sym^k(\IC) \equiv \IC^k/S_k$ (symmetric because permuting the
$z_i$ leaves the configuration unchanged; the solitons are like identical
particles). However, such symmetric products can be singular at the locus
of coincidence, and need to be looked at with care.

Let us consider next the case of $k=2$ and 
see what happens when two solitons are brought together. The description
of the corresponding projection operator in terms of the subspace spanned
by coherent states $\{|z_1\ra,|z_2\ra\}$ becomes bad since it seems as
if we have only one independent vector at the coincident point.
Actually, as $z_2$ approaches $z_1$, it is
not the subspace that becomes singular but simply our description of
it. Instead we should describe it as 
spanned, for example, by the vectors $|z_1\ra$ and
$(|z_2\ra-|z_1\ra)/(z_2-z_1)$. This basis has the virtue that it is
non-singular in the limit $z_2\to z_1$; in fact 
the second vector goes to $\p_{z_1}|z_1\ra=\adag|z_1\ra$. Now it is
clear that we are still in the moduli space in this limit, since the
limiting plane, spanned by $|z_1\ra$ and $\adag|z_1\ra$, is
obviously an invariant subspace of $a$.
It is straightforward to generalize this proof for any number $n$ of
solitons coming together. In appendix A we show that
\eqn\zlimit{
\lim_{z_i\to z}
{\rm span}\left\{|z_1\ra,\dots,|z_n\ra\right\} =
{\rm span}\left\{|z\ra,\adag|z\ra,\dots,(\adag)^{n-1}|z\ra\right\}.
}
Again, the right hand side is obviously an invariant subspace of $a$.

Conversely, one can see that
any finite dimensional invariant subspace of $a$ is a direct sum of spaces
of the form \zlimit, simply by writing the restriction of $a$ to the
subspace in Jordan form.
That the moduli space $\CM_k$ does not degenerate
in any way at the coincidence locus is a reflection
of the mathematical statement that $\Sym^k(\IC)$ is smooth everywhere. The
coordinates $z_i$ are bad coordinates near the coincidence locus but
in fact $\Sym^k(\IC)$ is isomorphic to $\IC^k$. 

Moreover, $\CM_k$ is isomorphic to $\IC^k$ as a complex manifold
since the spanning vectors $|z_i\ra$
depend holomorphically on the coordinates $z_i$. The isomorphism of
complex structures extends
to the coincidence locus since the change of basis matrix
used in appendix A to go to the non-singular description
also depends holomorphically on the $z_i$. This also implies that $\CM_k$
is a complex submanifold of $\Gr(k,\CH)$.

It is straightforward, using \pdef\ and the inverse Weyl-Moyal
transformation, to find the physical field configuration $\ph(w,\wb)$
corresponding to any given point in $\CM_k$. For example, with the
$z_i$ all distinct, we have
\eqn\phiform{
\ph(w,\wb) =
2\l\sum_{i,j=1}^kh_{ij}h^{ji}e^{-2(\wb/\sqrt{\t}-\zb_i)(w/\sqrt{\t}-z_j)}.
}
Figure 2 illustrates how the process of two solitons
coming together appears in the real space, while figure 3 shows an example
of a quadruple soliton. 

\fig{
The double soliton at various separations. Top: At large separation (here
$z=\pm2$), the solitons have the same
shape as single solitons \solone. Middle: As they come together they begin
to coalesce (here $z=\pm1$). Bottom: When
they coincide they form a rotationally symmetric ``level 2'' soliton.
The vertical axis is
$\phi(w,\wb)/\l$ and the horizontal plane is $w/\sqrt{\t}$.
}{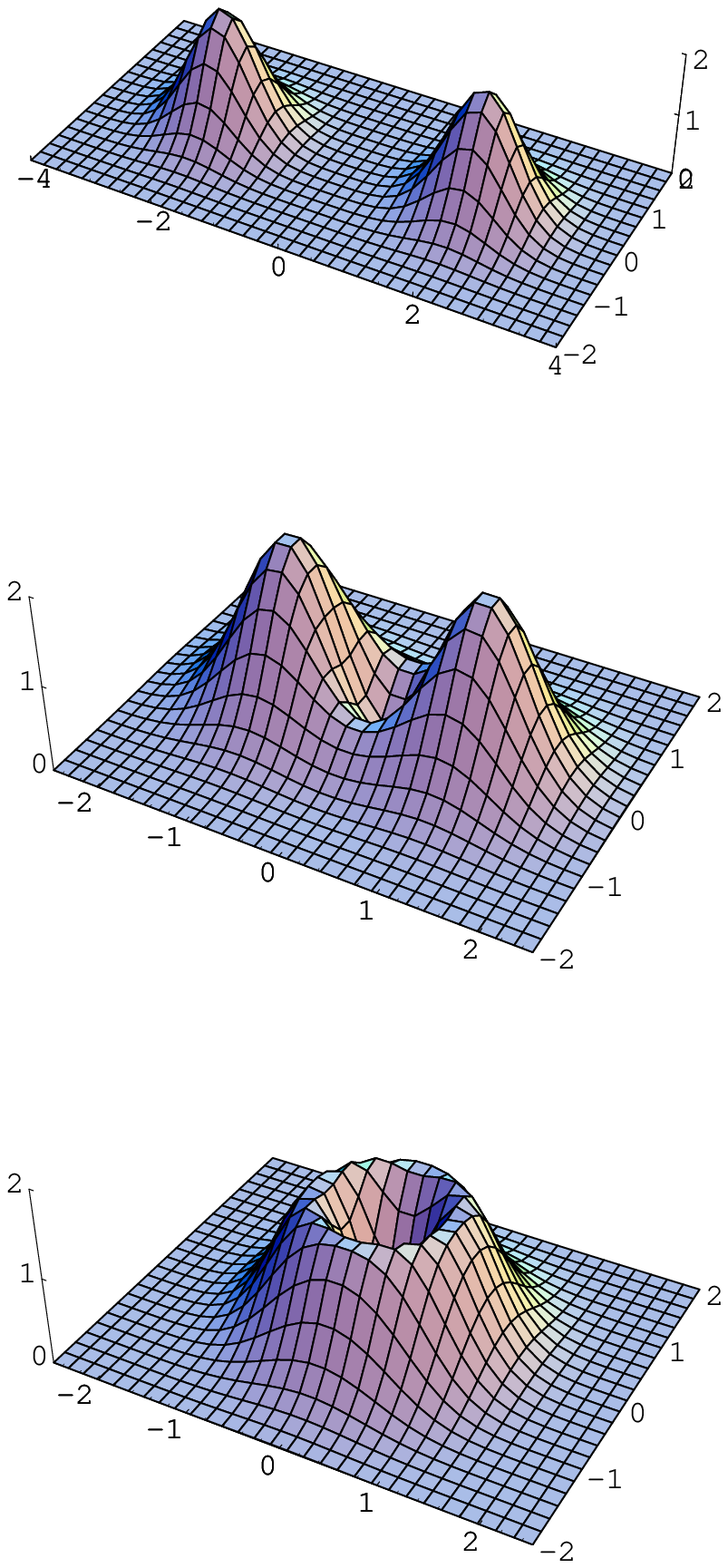}{2.7in}

\fig{
A quadruple soliton consisting of a triple soliton at the origin and a
single soliton at $z=2$. The axes are as in figure 2.
}{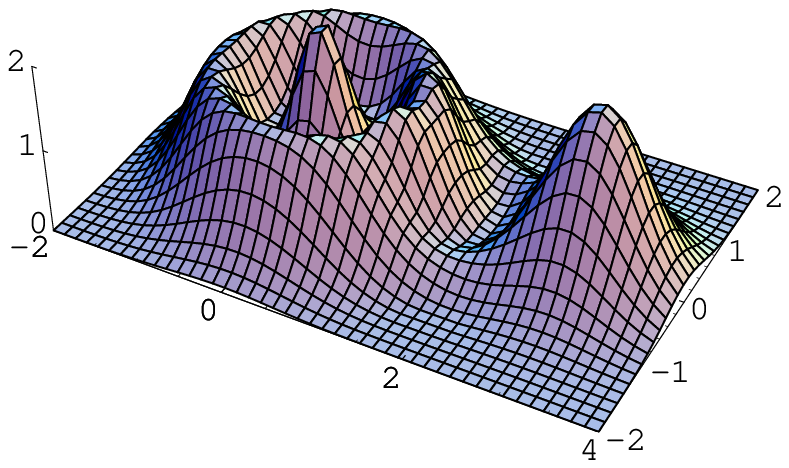}{3.5in}

\subsec{Geometry}

The flat metric $ds^2 = dz_id\zb_i$ on $\Sym^k(\IC)$ has conical
singularities on the coincidence locus. The solitons are not pointlike,
however, and the smoothness of the merging process depicted in figure 2
suggests that they somehow round out these singularities. 
Indeed, the moduli space $\CM_k$ has a smooth K\"ahler metric on it.
In this subsection we make some general observations about the
K\"ahler metric and investigate in detail the simplest case of $k=2$ (which
has appeared in \lru), but
the general proof of smoothness is relegated to appendix A.

The fact that $\CM_k$ is a complex submanifold of the K\"ahler manifold
$\Gr(k,\CH)$ implies that $\CM_k$ itself is K\"ahler. The K\"ahler
potential is given by \kahpot:
\eqn\kahler{
K = \ln\det\left(\la z_i|z_j\ra\right)
= \ln\det(e^{\zb_iz_j}).
}
(There is an overall factor of $2\pi\t\l^2$ in the physical
moduli space metric, which we will ignore
here.)

Two important features of the geometry are
immediate consequences of this form for K\"ahler potential.
Firstly, as the translational symmetry of the field theory implies, the
center of mass coordinate $c\equiv{1 \over k}
\sum_i z_i$ factors out, and the geometry
for this modulus is the flat plane:
\eqn\centmass{
K = k|c|^2 + \ln\det(e^{\bar y_iy_j}),
}
where $y_i=z_i-c$ are the relative coordinates. Secondly, when the
separations $|z_i-z_j|$ are large,
the determinant is dominated by the diagonal, so the metric
reduces to the flat one on $\Sym^k(\IC)$:
\eqn\Kapprox{
K \approx \sum_{i=1}^k|z_i|^2 \qquad (|z_i-z_j|\gg 1).
}

A third property
of the K\"ahler potential is also apparent from equation
\kahler: it diverges
on the coincidence locus. This is merely
a coordinate singularity.  The problem is the same as the one we
faced in the last subsection: the basis of coherent states $|z_i\ra$ is
degenerate when two or more of the $z_i$ coincide.
The solution is also the same: make a holomorphic change
of basis to a non-degenerate basis.
We show in appendix A that the
Jacobian of this change of basis is the Vandermonde
determinant $\prod_{i>j} (z_i-z_j)$,
and when the K\"ahler potential \kahler\ is calculated in this
basis one obtains
\eqn\regkahler{
K' = K - \sum_{i>j}\ln|z_i-z_j|^2.
}
The K\"ahler potentials
$K'$ and $K$ yield the same metric away from the coincidence
locus, but only the metric obtained from
the correct potential $K'$ extends smoothly to the locus.

Let us see in detail how this works in the case of $k=2$.
Since the center of mass coordinate plays no role, we will put
the two solitons at $z_1=z$ and $z_2=-z$ respectively. 
But $z$ is not a good coordinate in the neighborhood of $z=0$,
due to the identification $z\sim-z$. The good
coordinate is instead $\s=z^2$, in terms of which 
the metric is perfectly well behaved near $\s=0$:
\eqn\doublekah{
K' = \ln\left({2\sinh2|z|^2\over4|z|^2}\right) 
= {2\over3}|\s|^2 + \CO(|\s|^4),
}
so
\eqn\doublemet{
ds^2 = \left({2\over3}+\CO(|\s|^2)\right) d\s d\bar \s.
}
The space retains the symmetry of the cone, as well as its geometry far
from the vertex, but the vertex itself is rounded out.

The fact that the moduli space is curved implies that there are
velocity-dependent forces among the solitons.\foot{
{\bf Note added in revised version}:
In the paper \hlru\ (which appeared after the first version of this 
paper was submitted to the archive), the question was raised as to
whether there are bound-state wave-functions due to the curvature of the
moduli space. At least for the
case $k=2$, we may answer this question in the negative. By using a Weyl
transformation to bring the relative moduli space to the plane, it is
straightforward to show that the eigenvalue equation for the laplacian
admits no normalizable solutions. }
One can use the metric derived from
\regkahler\ to study the motion of these solitons on the plane, by
calculating geodesics on the moduli space \manton.
For instance, two-soliton scattering 
at zero impact parameter has been shown to lead to ninety degree 
deflection \lru . 
This can be understood as an immediate consequence
of the smoothness and rotational symmetry of the moduli space: A process
where two solitons start at
$z = \pm z_0$ and move toward each other is represented in the moduli
space by starting at the point $\s=z_0^2$ and moving toward the origin. 
The system
will pass smoothly through the origin, and by symmetry then pass through
the point $\s = -z_0^2$, representing the configuration 
where the solitons are at $z = \pm iz_0$.

For $k>2$ the
metric is complicated and therefore difficult to study, but one
interesting phenomenon which appears
is that the velocity-dependent forces tend to spread coincident
solitons apart when they move in the presence of other solitons. 
For example, if a
double soliton moves in the presence of a stationary third soliton, the
pair will spread apart along the direction of their motion.

\newsec{The effective potential on the moduli space}

The moduli space described in the previous section is not protected by any
symmetry and is therefore lifted, by classical and presumably quantum
effects.  A classical effective potential on the moduli space arises from
the fact that the solitons $\phh_0$ we have been discussing so far are not
exact solutions, but receive ${\cal O}(1/(\t m^2))$ corrections (the term
$\phh_1$ in \perturb). We will calculate these
corrections explicitly in this section and show that the effective
potential leads to an
attractive force among the solitons. As we will see, this potential, in
addition to being parametrically suppressed, is also well behaved. It is
bounded both from below and from above and hence is a small perturbation (for
large $\t m^2$) to the moduli space that we studied in the previous
section.

In the perturbation series \perturb\ for the energy, the first sub-leading
correction is
\eqn\etwo{
E_2[\phh_0,\phh_1] =
2\pi\,\Tr\left(
2[a,\phh_0][\phh_1,\adag] + {1\over2}\phh_1^2V''(\phh_0)
\right).
}
(We have eliminated the term involving $\phh_2$ due to the vanishing of
$V'(\phh_0)$.) Fixing $\phh_0$, the equation of motion for $\phh_1$
obtained from \etwo\ is easily solved:\foot{
We have ignored the issue of operator ordering, which is justified a
posteriori by the fact that (as one can readily verify) the solution
$\phh_1^{\rm c}$ commutes with $\phh_0$. There is in fact a continuous
family of solutions to the correct equation of motion
$2[\adag,[a,\phh_0]]+\left(\phh_1V''(\phh_0)\right)_{\rm s}=0$ (where the
subscript s indicates symmetrization over orderings of $\phh_1$ and
$\phh_0$), of which
$\phh_1^{\rm c}$ is the unique one that commutes with $\phh_0$.
The other solutions differ from $\phh_1^{\rm c}$ by an operator of the
form $PB(1-P)+(1-P)B^\dagger P$ for some $B$. One way to understand this
is that operators of this form perturb the solution $\phh_0$ in an
``angular'' direction (parallel to the infinite $\t$ moduli space), and
the second derivatives of $\Tr\,V$ vanish in those directions.
}
\eqn\phone{
\phh_1^{\rm c} = -2[\adag,[a,\phh_0]]V''(\phh_0)^{-1}.
}
Plugging this solution back into \etwo, we find, after some algebra, that
the correction to the energy is
\eqn\ecorrect{
V_{\rm eff}(P) \equiv 
{1\over\t m^2}E_2[\phh_0,\phh_1^{\rm c}(\phh_0)] =
-\CE\Tr\, G(P)^2,
}
where
\eqn\cedef{
\CE \equiv {4\pi\l^2\over\t m^2}\left(1 + {1\over V''(\l)}\right)
}
and
\eqn\Gdef{
G(P) \equiv
Pa(1-P)\adag P = P+[P\adag,aP].
}

The first way of writing $G(P)$ makes it obvious that it is non-negative
definite, while the second way makes it obvious that $\Tr\,G(P)=k$.
Together these two facts imply that $\Tr\,G(P)^2$ is bounded below by
$k$ and above by $k^2$, i.e.\ the effective potential is bounded above by
$-k\CE$ and below by $-k^2\CE$. The upper bound is achieved when
$G(P)=P$, i.e.\ $[P\adag,aP]=0$; this
occurs only in the limit that the solitons are infinitely far from each
other, since only in this limit is $aP$ unitarily diagonalizable.

The lower bound, on the other hand, is achieved when
$G(P)$ has only a single nonzero eigenvalue, equal to $k$. Let the
corresponding eigenvector be $|\psi\ra$, so that
$G(P)=k|\psi\ra\la\psi|$.
Then for every $|\psi'\ra$ orthogonal to $|\psi\ra$,
\eqn\symcond{
(1-P)\adag P|\psi'\ra=0,
}
which in turn implies that the image of $P$ is of the form of the
right-hand side of \zlimit. (Otherwise, if the image of $P$ were the
direct sum of two or more spaces of this form, then every linear
combination of
the states $(\adag)^{n_1-1}|z_1\ra$ and $(\adag)^{n_2-1}|z_2\ra$ would
violate \symcond.  Then $G(P)$ would not be of the desired form because
it would have two nonzero eigenvalues.) So the
effective potential is minimized 
by configurations in which all $k$ solitons are coincident. The moduli
space of such configurations is obviously just the plane itself. In other
words, all the
moduli $y_i$ for relative motion are lifted. The remaining center-of-mass
modulus $c$ is exact by the translational symmetry of the theory, and
cannot be lifted at any order in $1/(\t m^2)$, or by quantum
corrections.

The two-soliton effective potential, in terms of the coordinate $\s =
(z_1-z_2)^2/4$, is
\eqn\twosolcor{
V_{\rm eff}(\s) = 
- 2\CE\left(1 + {4 |\s|^2 \over \sinh^2(2|\s|)}\right),
}
which is plotted in figure 4. The force is attractive, but falls off
exponentially when the solitons are more than a few multiples of
$\sqrt{\t}$ apart. At higher $k$, the functional form of $V_{\rm eff}$ is
more complicated, but it retains these features.

\fig{
The two-soliton effective potential \twosolcor: $V_{\rm eff}/\CE$
versus $|\s|=|z_1-z_2|^2/4$.
}{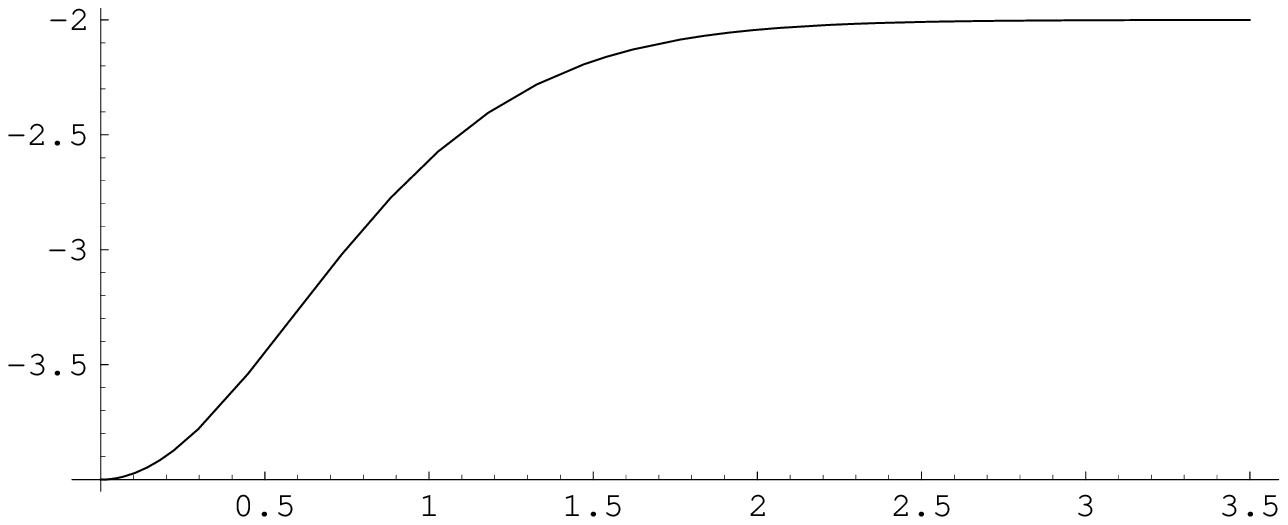}{3.9in}

\newsec{Solitons on quotient spaces}

We can exploit the results of section 3 for multi-solitons to construct
solitons on quotients of $\IR^2$ by various discrete symmetry groups. The
basic principle is simple:
construct a multi-soliton on the covering space that respects the
quotienting symmetry.

\subsec{$\IR^2/\IZ_k$}

As a simple example, consider the orbifold $\IR^2/\IZ_k$. We may put $k$
solitons at the vertices of a regular polygon: $z_i = \w^i z_0$, where $\w
\equiv e^{2\pi i/k}$. Such configurations form a submanifold of $\CM_k$,
parametrized by the single modulus $z_0$; by the rotational symmetry of
the theory this submanifold is totally geodesic. This 2-dimensional moduli
space is the same orbifold $\IR^2/\IZ_k$, since $z_0$ is identified with
$\w z_0$. Geometrically, however, its K\"ahler structure is deformed, with
the conical singularity at the orbifold fixed point smoothed out, as
described in subsection 3.3. This is an
example of the stringy behavior of noncommutative solitons: as non-local
probes, they see a singular geometry in a non-singular way.
The attractive potential of order $\l^2/(\t m^2)$ between the soliton and
its images, described in section 4,
will draw it toward the fixed point.

\subsec{Cylinder}

A soliton on a cylinder of circumference $\sqrt{\t}l$ can be represented 
as an infinite array of solitons on the plane, located at 
$z_j = z_0 + ijl/\sqrt{2}$, $j\in\IZ$, where
$z_0$ is the modulus. Since the moduli space is quotiented under the
identification $z_0 \sim z_0 + il/\sqrt{2}$, it has the same geometry as
the underlying cylinder. Furthermore, this moduli space is exact by the
translational symmetry of the theory---
the attraction between the soliton and its images on one side balance the
attraction of those on the other side.

In finding the explicit field configuration
$\ph(w,\wb)$ of this soliton, equation \phiform\ is not of much use, since it
requires inverting an infinite dimensional matrix. We will therefore
present an alternative construction.      
For simplicity we will assume in what
follows that $z_0 = 0$. It will be convenient to use real coordinates $y^a
\equiv x^a/\sqrt{\t}$ (recall that $w = (x^1+ix^2)/\sqrt{2}$), and the
corresponding hermitian operators $\hat y^a$. The discrete translations by
which the plane is quotiented are generated by the unitary operator
$U\equiv e^{il\hat y^1}$. We thus wish to construct the projection
operator whose image is spanned by the set $\{U^j|0\ra:j\in\IZ\}$ of
(normalized) coherent states. To do this, 
we will find an orthonormal basis for this
hyperplane of the form $\{U^j|\psi\ra:j\in\IZ\}$, where
\eqn\orthonorm{
\la\psi|U^j|\psi\ra = \delta_{j0}.
}
This will allow us to decompose $P$ as a sum of orthogonal
projection operators:
\eqn\Pdecomp{
P = \sum_j U^j|\psi\ra\la\psi|U^{-j}.
}
The inverse Weyl-Moyal transformation then yields
\eqn\phiformcyl{
\ph(y^1,y^2) =
{2\pi\l\over l}\sum_j
e^{-2\pi ijy^2/l}
\psi^*(y^1-{\pi j\over l})\psi(y^1+{\pi j\over l}),
}
where $\psi(y^1)\equiv\la y^1|\psi\ra$ is the wave function of $|\psi\ra$
in the basis of $\hat y^1$ eigenstates.

We will find the vector $|\psi\ra$ by finding its coefficients in the
$U^j|0\ra$ basis, or more precisely, the Fourier transform of those
coefficients:
\eqn\psiexp{
|\psi\ra = \sum_j c_jU^j|0\ra = \tilde c(l\hat y^1)|0\ra.
}
Working in terms of the wave functions, with
\eqn\wavfun{
\psi_0(y^1) \equiv \la y^1|0\ra = \pi^{-1/4}e^{-(y^1)^2/2},
}
we impose the orthonormality condition \orthonorm,
\eqn\findc{\eqalign{
\delta_{j'0} =&
\int dy^1\,
e^{ij'ly^1}|\tilde c(ly^1)|^2\psi_0(y^1)^2
\cr
=&
\int_0^{2\pi/l}dy^1\,
e^{ij'ly^1}|\tilde c(ly^1)|^2
\sum_{j}\psi_0({\textstyle{y^1+{2\pi j\over l}}})^2.
}}
We can now solve for $\tilde c$ (choosing it to be real) and thus for $\psi$ :
\eqn\psidef{\eqalign{
\psi(y^1)&= 
\sqrt{{l\over2\pi\sum_j|\psi_0(y^1+2\pi j/l)|^2}}\psi_0(y^1) \cr
&= \sqrt{{l\over2\pi\vartheta_{00}(2iy^1/l,\tau)}},
}}
where $\tau \equiv 4\pi i/l^2$.\foot{
In fact, this construction 
is a very general one, applicable to
constructing an arbitrary soliton on the cylinder at $\t=\infty$. In
other words we can construct arbitrary projection operators of the form
\Pdecomp\ which respect the discrete translation symmetry. All one needs
to do is replace $|0\ra$ in \psiexp\ and \wavfun\
by an arbitrary ket $|\tilde\psi\ra$. The solution for the coefficients
$\tilde c(l\hat y^1)$ is then as in the first line of \psidef. Examples of
such general
solitons on the cylinder have also been constructed 
by V. Balasubramaniam and A. Naqvi
(unpublished).} 
Plugging this into \phiformcyl, we find
\eqn\field{
\ph(y^1,y^2) =
\l\left(
{\vartheta_{00}(2y^2/l,\tau)\over\vartheta_{00}(2iy^1/l,\tau)}
+{\vartheta_{10}(2y^2/l,\tau)\over\vartheta_{10}(2iy^1/l,\tau)}
\right).
}
This is plotted in figure 5 for various values of $l$. One can show
by a modular transformation that in the decompactification limit
$l\to\infty$, \field\ goes over to the single soliton on the plane.

\fig{
The cylinder soliton \field\ at various 
circumferences: $l=3$ (top), $l=2$ (middle), and $l=1$ (bottom). One and a
half copies of the soliton are displayed. The axes are as in figure 2.
}{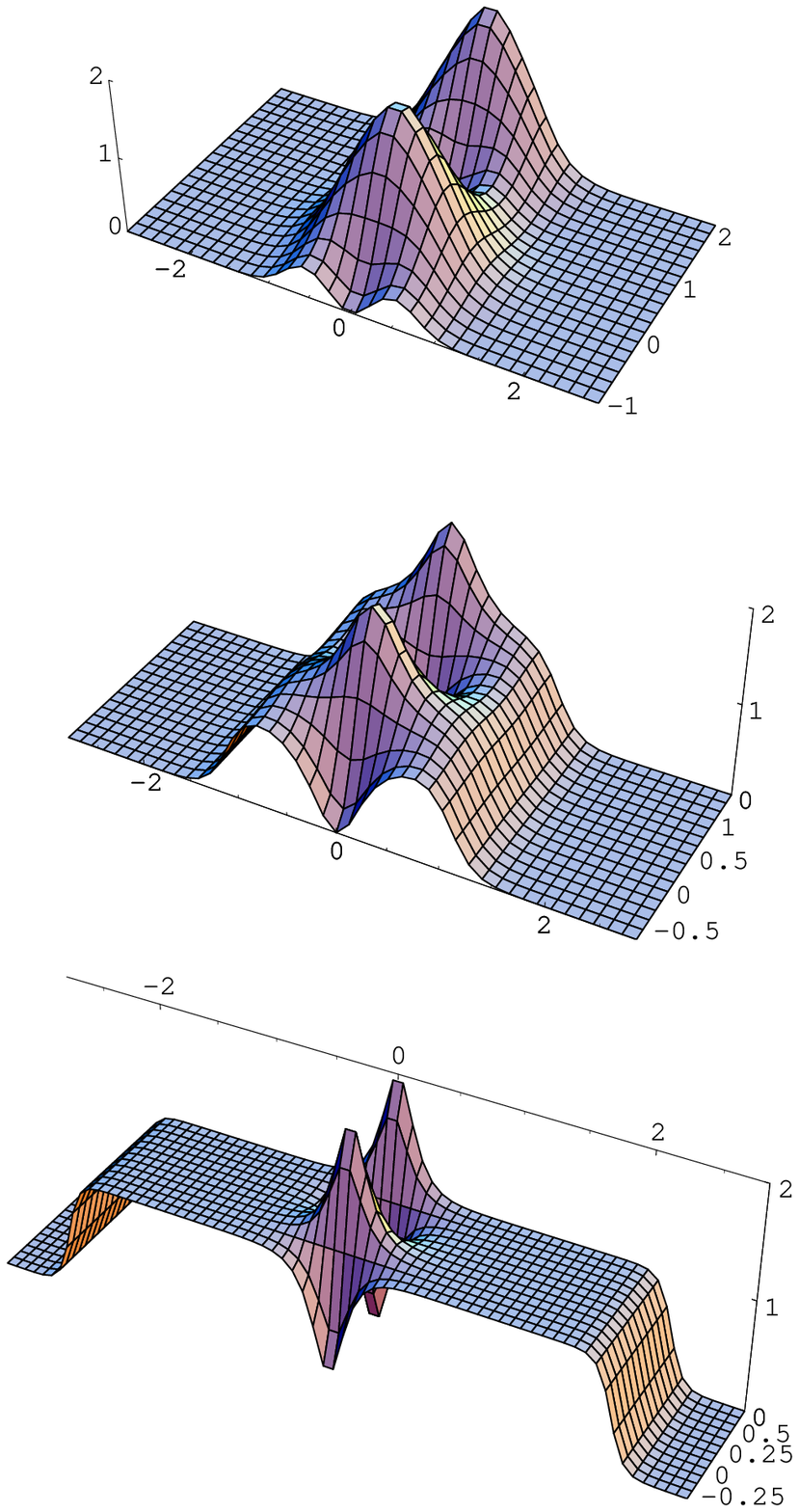}{3.4in}

\subsec{$T^2$}

Just as in the case of the cylinder, a soliton on the torus can be
represented as a lattice of solitons on the plane. With $\tau$ the modular
parameter of the torus and $\sqrt{\t}l$ its circumference in the $x^1$
direction,
place the solitons in the $z$-plane at $z_{j_1j_2} = (j_1 +
j_2\tau)l/\sqrt{2}$, $j_1,j_2\in\IZ$. Again, the forces between the
soliton and its images will balance out, and the moduli space (which is
the torus itself) is exact.

In principle there is no
problem defining the projection operator whose image is the
infinite dimensional hyperplane spanned by
$\{|z_{j_1j_2}\ra:j_1,j_2\in\IZ\}$, for any size and shape of torus. In
practice, however, we have only been able to
find an explicit expression for the field configuration corresponding to
such a projection operator in
the special case where the parameter $A \equiv \tau_2l^2/2\pi$ (which
is area of the torus in units of $2\pi\t$)
is an integer. A major simplification occurs in this case because then the
operators $U_1 \equiv e^{-il\hat y^2}$ and 
$U_2 \equiv e^{il(\tau_2\hat y^1-\tau_1\hat y^2)}$
generating the two lattice translations commute, and the same
method as that used for the cylinder above can be employed.
The details of the derivation are presented in appendix B; the final
result is:
\eqn\torus{\eqalign{
\ph(w,\wb) = 
{\l\over2}&\left(
\frac{\sum\vartheta^*_{00}(\nu-{n\over A})\vartheta_{00}(\nu+{n\over A})}
{\sum\left|\vartheta_{00}(\nu+{n\over A})\right|^2}+
\frac{\sum\vartheta^*_{00}(\nu-{n+1/2\over A})
\vartheta_{00}(\nu+\frac{n+1/2}{A})}
{\sum\left|\vartheta_{00}(\nu+\frac{n+1/2}{A})\right|^2} \right. \cr
&\left.{}+\frac{\sum\vartheta^*_{10}(\nu-\frac{n}{A})
\vartheta_{10}(\nu+\frac{n}{A})}
{\sum\left|\vartheta_{10}(\nu+\frac{n}{A})\right|^2}+
\frac{\sum\vartheta^*_{10}(\nu-\frac{n+1/2}{A})
\vartheta_{10}(\nu+\frac{n+1/2}{A})}
{\sum\left|\vartheta_{10}(\nu+\frac{n+1/2}{A})\right|^2}
\right),
}}
where $\nu \equiv \sqrt{2}w/(\sqrt{\t}l)$, all theta functions take
$\tau/A$ as their second argument, and all sums run from $n=0$ to $A-1$.
It is not obvious how to generalize such a formula to the case where $A$
is not an integer. The shape of this soliton for the case $A=2$ is plotted
in figure 6.

\fig{
The soliton \torus\ on the torus of area $4\pi\t$ ($A=2$) for two values
of the modular parameter: the square torus $\tau=i$ (top), and
$\tau=3i+1/2$
(bottom). Four
copies of the torus are displayed. The axes are as in figure 2.
}{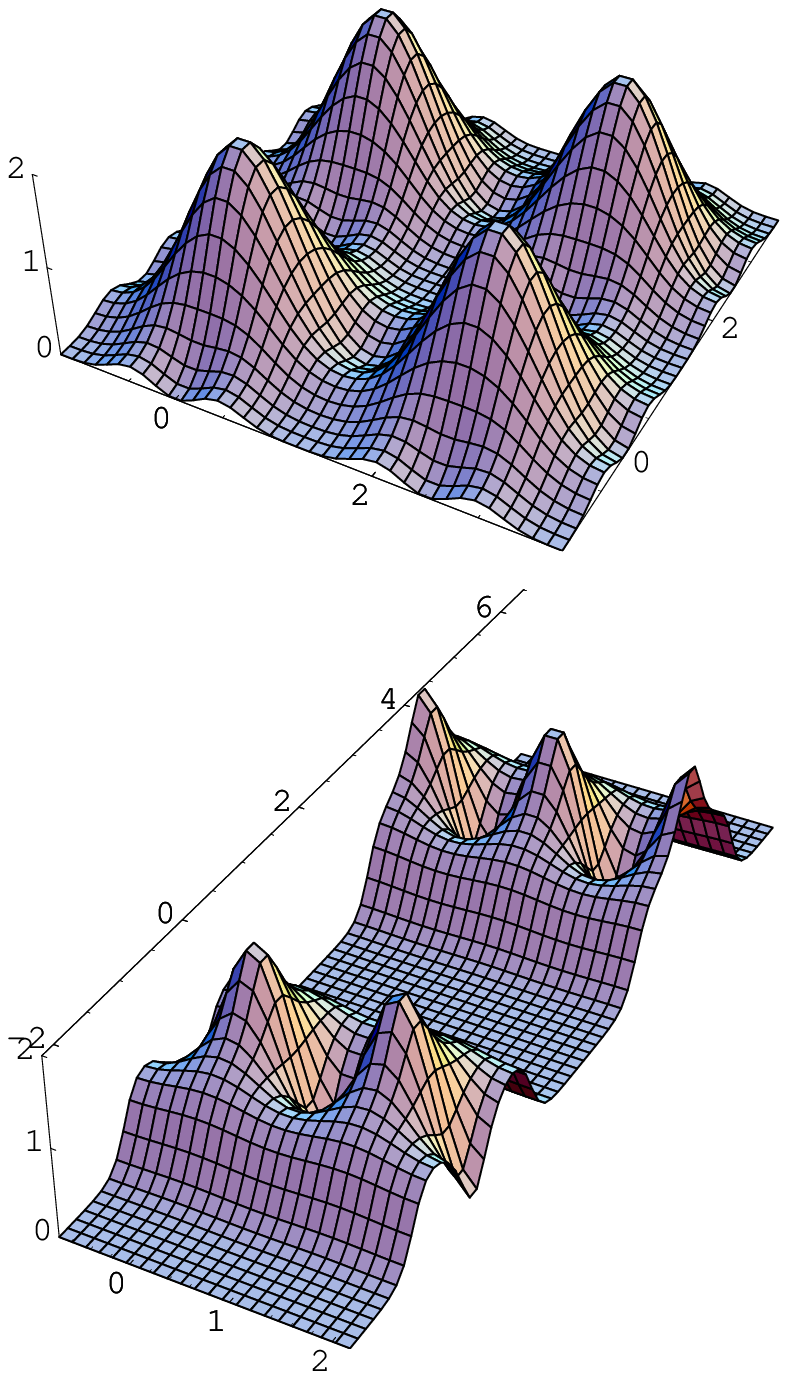}{2.7in}

Consider now the alternative case where $1/A$ is an integer. Here the
Moyal star product is in fact equivalent to ordinary pointwise
multiplication of
functions, and so a small puzzle arises: No non-trivial continuous
solutions to the equation of motion $\ph\star\ph=\l\ph$ are possible, yet
one can
certainly define the projection operator $P$ whose image is the hyperplane
spanned by $\{|z_{j_1j_2}\ra:j_1,j_2\in\IZ\}$. It is not hard to guess the
resolution: this lattice of coherent states actually spans
the entire Hilbert space, so $P$ is just
the identity, and the ``soliton'' is the constant solution $\ph = \l$.
Indeed, the formula \torus\ shows explicitly that this is the case when
$A=1$. But we can say even more: it is known
\refs{\perelomov-\bgz}
that any lattice of coherent states with $A\le1$ is a
complete (actually, overcomplete) basis for the
Hilbert space.\foot{
The case $A=1$ is known as a von Neumann lattice, since it was first
discussed by him in \vn, where the claim was made (without proof) that
such a lattice forms a complete basis.
An amusing fact about von Neumann lattices is that they are
overcomplete by exactly one state, that is, there is exactly one linear
relation among the infinitely many states. Lattices with $A<1$ are
overcomplete by infinitely many states.
} We thus conclude that there are no stable scalar solitons on any
torus---rational or irrational---of area less than or equal to $2\pi\t$.

\newsec{Solitons in higher dimensions}

The analysis of this paper easily generalizes to $2d+1$ dimensional
scalar field theory with spacelike noncommutativity, although we
will see that the topological structure of the moduli space $\CM_k^d$ in
higher dimensions is much richer than in $d=1$.

We choose complex coordinates $w_r$, $r=1,\ldots,d$,
in which the noncommutativity
parameter takes the form $\t^{r \bar{s}} = - i \delta_{rs}
\t_s$.
The Weyl-Moyal correspondence maps the field
$\phi$ to an operator $\phh$ on the Hilbert space
$\CH^d \equiv \CH \otimes \cdots \otimes \CH$
of a particle in $d$ dimensions, with
$w_r \to \sqrt{\t_r} a_r$ and
$\p_{r} \to -{1 \over \sqrt{\t_r}} [\adag_r,\cdot\,]$.
The moduli space at infinite\foot{
By this we mean the limit in which all of the
$\t_r$ are taken to infinity with their ratios held fixed, while
the energy is rescaled
by a factor of $m^2 \t_1 \cdots \t_d$, generalizing
\ezero.} $\t$
again consists of operators of the form $\phh = \l P$ for
any projection operator $P$ of a given rank $k$.
The Bogomolnyi bound \bogom\ on the leading contribution from the
kinetic energy takes the form
\eqn\bogomd{\eqalign{
E_1 [ \phh_0] &= (2 \pi)^d  \l^2 \t \sum_{r=1}^d
{1 \over \t_r}\Tr [a_r,P][P,\adag_r]\cr
&= (2 \pi)^d  \l^2
\t
\sum_{r=1}^d {1 \over \t_r}
\Tr \left(  P + 2 F_r(P)^\dagger
F_r(P) \right)\cr
&\ge (2 \pi)^d  \l^2 \t k \sum_{r=1}^d {1 \over \t_r},
}}
where $\t \equiv \t_1 \cdots \t_d$ and
$F_r(P) \equiv (1-P) a_r P$.
The moduli space $\CM_k^d$ at finite $\t$ 
is therefore the space of projection operators on $\CH^d$ whose image
is an invariant subspace of  all of the $a_r$.

A large class of such operators may be constructed
by letting $P$ project onto the space spanned by $k$
independent
coherent states  $|\vec{z}_i \ra \equiv e^{\vec{z}_i \cdot
\vec{a}^\dagger}
|\vec{0}\ra$.
Such a $P$ corresponds to a multi-centered gaussian with peaks
at $\vec{z}_i$.
Naively, the moduli space of such
solitons seems to be $\Sym^k(\IC^d)$, as we saw is
indeed
the case in $d=1$.
However, in dimension $d>1$ coincident solitons have
more moduli than are present in the symmetric product, and
we will see that the
full moduli space $\CM_k^d$ is the so-called
Hilbert scheme $\Hilb^k(\IC^d)$
of $k$ points in $\IC^d$.
Before introducing the general machinery of Hilbert
schemes, we turn in the next subsection to the relatively
simple case of $d=2$
in order to gain some insight into the geometry of the moduli
space when higher dimensional solitons come together.

\subsec{$4+1$ dimensions}

Consider first the case $k=2$, with two separated solitons
described by the projection operator onto the space
spanned by $|\vec{z}_1\ra$ and $| \vec{z}_2 \ra$.
When the two solitons come together we have
\eqn\twolim{
\lim_{\vec{z}_i \to \vec{z}} {\rm span} \{ | \vec{z}_1 \ra,
| \vec{z}_2 \ra \} =
 {\rm span} \left\{ |\vec{z}\ra, \vec{\gamma} \cdot \vec{a}^\dagger
|\vec{z}\ra \right\}, \qquad {\rm where}~\vec{\gamma} \equiv
\lim_{\vec{z}_i \to \vec{z}} {\vec{z}_1 -\vec{z}_2 \over |\vec{z}_1-
\vec{z}_2|}.
}
Thus the ``origin'' of the relative moduli space is not a single point,
but rather a $\IP^1$ parametrized by the complex direction
$\vec{\gamma}$ along which the two solitons came together.
This is in contrast to the $d=1$ case studied in section 3, where
we found that two solitons brought together along any direction
end up at the unique point span$\{ |z\ra, \adag |z \ra \}$.

The physical significance of the sphere hiding at the coincidence locus
is made clear by applying the Weyl-Moyal correspondence to the solution 
$\phh_{\vec \gamma} = \lambda P(\vec{\gamma})$, where $P(\vec{\gamma})$ is
the projection operator onto 
span$\{|\vec{0}\ra,\vec{\gamma} \cdot \vec{a}^\dagger|\vec{0}\ra\}$.
This yields the function
\eqn\aaa{
\phi_{\vec{\gamma}}(\vec{w})= 16 \lambda
e^{-2 |w_1|^2/\t_1 - 2 |w_2|^2/\t_2} { |w_1 \gamma_1/\sqrt{\t_1}
+ w_2 \gamma_2/\sqrt{\t_2}|^2 \over |\vec{\gamma}|^2},
}
from which we see that
the modulus $\vec{\gamma}$ encodes information about the shape
of the level two lump.  In other words, whereas two coincident
solitons in $d=1$ have
(other than the overall translational mode)
only a modulus corresponding to separating
the two solitons,
in $d=2$ there is in addition a modulus
corresponding to deforming the lump.

Factoring out the $\IC^2$ center of mass, we can parametrize
the relative moduli space with coordinates $z \in \IC$ and
$\vec{\gamma} \in
\IC^2\setminus \{\vec{0}\}$ by letting
\eqn\ktwopdef{
{\rm im}\,P(z,\vec{\gamma}) = {\rm span} \{| z \vec{\gamma}
\ra, |{-z\vec{\gamma}}\ra \}.
}
These coordinates are subject to the identification $(z,\vec{\gamma})
\sim (z/\l, \l \vec{\gamma})$ for $\l \in \IC
\setminus \{0\}$, which 
we recognize as
defining the complex line bundle $\CO(-1)$ over $\IP^1$.
The base $\IP^1$ sits at $z = 0$, which can be seen from
$\lim_{z \to 0} {\rm im}\, P(z,\vec{\gamma}) = P(\vec{\gamma})$
as defined above.

Plugging \ktwopdef\ into \kahpot\ gives
the K\"ahler potential on the moduli space
\eqn\ktwopot{
K' = \ln\left({2 \sinh \left( 2 |z|^2 |\vec{\gamma}|^2 \right) \over
 |z|^2}\right),
}
where the factor in the denominator arises as in \regkahler\ from the
Jacobian of  the transformation to a basis which remains nondegenerate
as $z\to 0$.
Note that the K\"ahler potential \ktwopot\ induces the familiar
Fubini-Study
metric on the $\IP^1$ at $z=0$.

We conclude that two solitons in $4+1$ dimensions resolve
the singular configuration space $\Sym^2(\IC^2)$ into a smooth complex
manifold with topology $\IC^2 \times \CO_{\IP^1}(-1)$
and a smooth K\"ahler metric given by \ktwopot.
In particular, the $\IP^1$ at the origin has finite area
$(2 \pi)^3 \l^2 \t_1 \t_2$.

For $k>2$ solitons in $4+1$ dimensions the topology of the
the moduli space is more complicated.
For example, for $k=3$ the relative moduli space has,
in addition to the $(\IC^2\setminus\{\vec{0}\}) \times \CO_{\IP^1}(-1)$
when any two solitons come together, a $\IP^1$
nontrivally fibered over $\IP^1$ at the ``origin'' of the relative
moduli space where all three solitons merge.
These moduli again correspond to deformations of the
shape of the soliton, in contrast to the situation in $d=1$
where the only moduli (other than overall translations)
of a collection of coincident solitons
correspond to separating some of the solitons from each other.

As we will see in the next subsection, $\CM_k^2$
has the same topology as the
moduli space of $k$ U$(1)$ instantons on noncommutative
$\IR^4$ \nekrasov, which is a smooth complex manifold for any $k$.
However, two important differences with that case warrant
mention.  Firstly, the metric on the scalar soliton moduli
space $\CM_k^2$ is only K\"ahler and not hyperk\"ahler.  
Secondly, nothing in our analysis requires that the noncommutativity
parameter $\t$ in $4$ dimensions be self-dual.
Indeed, the precise values of $\t_1$ and $\t_2$ scale out
of the problem---their only role is to set the scale of the
physical coordinates $w_r$ with respect to the dimensionless
moduli space coordinates $z_i$.  The moduli space
thus possesses an accidental SU$(2)$
symmetry which is not possessed by the full theory when
$\t_1 \ne \t_2$.

\subsec{Hilbert schemes of points}

In this subsection we unify the discussion of the moduli space
of $k$ solitons in $2d + 1$ dimensions by showing
that $\CM_k^d$ is isomorphic to a mathematical object known
as the Hilbert scheme $\Hilb^k(\IC^d)$ of $k$ points in $\IC^d$.

$\Hilb^k(\IC^d)$
is defined as the set of ideals $\CI$ of codimension $k$ in the polynomial
ring $\IC[x_1,\ldots,x_d]$.
The correspondence between projection operators and ideals
is intuitively clear: if $f$ is a polynomial in some ideal $\CI$
and $g$ is any polynomial, then the polynomial $fg$ is still in $\CI$.
Therefore the polynomials in an ideal $\CI$ may be thought of roughly
as projection
operators from all of $\IC[x_1,\ldots,x_d]$ into $\CI$.
The precise correspondence we demonstrate is motivated
by a nearly identical correspondence
between projection operators
on a Fock space
and ideals in polynomial rings
(see for example \refs{\nekrasov, \furuuchi})
which appears in the construction of noncommutative instantons
on $\IR^4$.

For any polynomial $f \in \IC[x_1,\ldots,x_d]$, we may construct the ket
\eqn\statedef{
|f \ra \equiv f(\adag_1,\ldots,\adag_d) | \vec{0} \ra \in \CH^d.
}
The one-to-one correspondence between $\Hilb^k(\IC^d)$
and the moduli space $\CM_k^d$ goes as follows.
For an ideal $\CI \subset \IC[x_1,\ldots,x_d]$, we let $1-P$ be
the projection operator onto (the closure of) the
linear subspace spanned by $\{ | f \ra : f \in \CI\}$.
Conversely, given $P \in \CM_k^d$,
we may recover $\CI = \{ f \in \IC[x_1,\ldots,x_d]
: P |f \ra = 0 \}$.  This $\CI$ is indeed an ideal by virtue of the
fact that $P \adag_r = P \adag_r P$.

For example, if $P$ projects onto the subspace spanned
by $k$ independent coherent states $|\vec{z}_i\ra$, then the corresponding
ideal is simply the set of all polynomials which vanish at the $k$ points
$\vec{z}_i$.
If $n$ of the points $\vec{z}_i$ come together at $\vec{z}$, then
the ideal becomes the set of polynomials with an order $n$ zero at
$\vec{z}$.

For $d=1$ we have simply $\Hilb^k(\IC) \cong \IC^k$, in agreement with
the result
of section 3.  For $k=2$ but arbitrary $d$, a trivial generalization
of the argument in the previous subsection shows that the origin
of the relative moduli space is not
just a point but a whole $\IP^{d-1}$.
The total space $\CM_2^d$ in this case is the center of
mass $\IC^d$ times the complex line bundle $\CO(-1)$ over
$\IP^{d-1}$.
For $d=2$ but arbitrary $k$, the Hilbert scheme $\Hilb^k(\IC^2)$
is a smooth
manifold of complicated topology which arises as the moduli
space of $k$ U$(1)$ instantons on noncommutative $\IR^4$
\refs{\nakajima-\bn}.

For $k>3$ and $d>2$, however, the Hilbert scheme is not smooth, and
in fact it is not even a manifold, having in general
several different branches of varying dimension.
We present an example in the next subsection.

\subsec{An exotic example}

It is not difficult to construct exotic branches of the moduli space
$\CM_k^d = \Hilb^k(\IC^d)$ for sufficiently large $k$ and $d$.
Let us consider $k=12$ solitons in $d=8$.\foot{
A nearly identical construction works in many other cases---in $d=3$,
for example, with as few as $k=97$ solitons.
}
If the moduli space $\CM_{12}^8$ only contained information about
the location of 12 solitons free to move about in $\IC^8$, then we
would expect it to be 96-dimensional.  However, we will now construct
a 99-dimensional submanifold of $\CM_{12}^8$ which opens up when all
of the solitons coincide.

We work in a harmonic oscillator
basis $\CH^8 = {\rm span} \{|n_1,\ldots,n_8\ra :
n_r \ge 0 \}$.
Let us denote by $|\alpha\ra$, $\alpha=1,\ldots,36$, the 36 basis
vectors which have $\sum n_r = 2$.
Then for any three linearly independent 36-component vectors
$w_1^\alpha$, $w_2^\alpha$ and $w_3^\alpha$, we can define the projection
operator $P$ by
\eqn\strangep{
{\rm im}\,P(w_1,w_2,w_3) = {\rm span}\{ |\vec{0}\ra,
|1,\ldots,0\ra, \cdots, |0,\ldots,1\ra, w_1^\alpha |\alpha\ra,
w_2^\alpha|\alpha\ra, w_3^\alpha|\alpha\ra \},
}
with summation over the $\alpha$ indices implied.
This operator projects onto a 12-dimensional subspace
of $\CH^8$ which is invariant under all of the $a_r$.
Therefore $P \in \CM_{12}^8$ for any three vectors $w$.  These
vectors
represent the choice of a three dimensional subspace of the 36-dimensional
space spanned by $|\alpha\ra$, and hence they parametrize
the Grassmannian $\Gr(3,\IC^{36})$.
Now $\Gr(3,\IC^{36})$ is 99-dimensional,
so \strangep\ gives a 99 parameter family of projection
operators inside of $\CM_{12}^8$.

The branch parametrized by projection operators of the form \strangep\ 
cannot be smoothly connected to the 96-dimensional branch where all of
the solitons are separated.  The analysis of the global structure of the
$\Hilb^k(\IC^d)$ for general $k>3$ and $d>2$,
including details about how the various branches connect
to each other, is a problem that mathematicians have only begun to tackle.

\vskip 0.5in

\centerline{\bf Acknowledgements}

\vskip .1in

It is a pleasure to thank
R. Glauber, T. Graber, D. Gross,
K. Hori, S. Katz, D. Khosla, A. Mikhailov, S. Minwalla, A. Naqvi, 
N. Nekrasov, K. Oh, B. Pioline, 
S. Sethi,
A. Strominger, L. Thorlacius, C. Vafa,
A. Volovich, M. Wijnholt, and S. Zhukov for helpful conversations.
We are especially grateful to S. Minwalla for comments on the
manuscript.
This work was supported in part by DOE grant DE-FG02-91ER40654 and the
David and Lucile
Packard Foundation. One of us (R.G.) would like to thank the I.T.P. for
hospitality and acknowledge the support of NSF grant PHY99-07949.

\appendix{A}{Completeness and smoothness of the finite $\t$ moduli space}

In this appendix we prove the various statements made in section 3
about the topology and geometry of the moduli space $\CM_k$
when $n$ points $z_a$
come together at a point $z$, with the other $k-n$ points fixed.
Note that throughout this appendix we will use indices $a,b =
1,\ldots,n$ while the indices $i,j$ will continue to
 run from 1 to $k$, as in the body of the paper.

The completeness of the moduli space follows
trivially from the Hilbert scheme analysis of section 6, but
nevertheless it is useful to see explicitly how to construct
a basis of states which is nondegenerate as two or more solitons
come together.  In particular, this is necessary to prove the
assertion made above \regkahler\ that the Jacobian of the required change
of basis is the Vandermonde determinant, and to verify that the
metric obtained from \regkahler\ extends smoothly to the coincidence
locus.

\subsec{Some elementary machinery}

The proofs are completely straightforward but require some rather heavy
notation, which we now introduce.
Define $u_a = z_a-z$.
We will make use of the $n \times n$ Vandermonde
matrix $V$ with entries
\eqn\vandermond{
V^b{}_a = u_a^{b-1},
}
which has determinant
\eqn\detv{
\Delta \equiv \det V = \prod_{a>b} (u_a-u_b).
}
We will also make use of the polynomials defined by
\eqn\fdef{
F_{ \{c_a\}}(u_a) = 
  \left| \matrix{
u_1^{c_1} & \cdots & u_n^{c_1} \cr
\vdots & & \vdots \cr
u_1^{c_n} & \cdots & u_n^{c_n}
}\right|,
}
where $\{c_a\}$ is a set of $n$ ordered nonnegative integers.
It is clear that $F_{\{c_i\}}$ is a homogeneous polynomial
in the $u_i$ which vanishes unless the $c_i$ are distinct.
The lowest degree non-zero one is therefore just the Vandermonde
determinant:
\eqn\blah{
F_{0,\ldots,n-1}(u_a) = \Delta.
}
Since the polynomial $F_{\{c_a\}}$  vanishes if any two of the $u_a$ are
equal, it must be divisible by $u_a - u_b$ for all $a$ and $b$, and
hence by $\Delta$. Therefore
\eqn\qdef{
Q_{\{c_a\}}(u_a) \equiv F_{\{c_a\}}(u_a)/\Delta
}
is again a homogeneous polynomial.  Furthermore, since both $\Delta$
and $F_{\{c_a\}}$ are antisymmetric under the interchange of any two
of the $u_a$, $Q_{\{c_a\}}$ must be a symmetric polynomial
in the $u_a$.
Now introduce an auxiliary variable $u$ and consider the
determinant
\eqn\cooldet{
\Delta \prod_{a=1}^n (u+u_a)
=
 \left| \matrix{
1 & 1 & \cdots & 1  \cr
-u & u_1 & \cdots & u_n  \cr
\vdots & \vdots & & \vdots\cr
(-u)^n & u_1^n & \cdots & u_n^n
}\right| = \sum_{b=0}^n u^b
F_{0,\ldots,\widehat{b},\ldots,n}(u_a)
}
where the notation $\widehat{b}$ means that the index $b$ is
to be omitted.
Upon dividing both sides of \cooldet\ by $\Delta$ we learn
that the $Q_{\{c_a\}}$ of degree $\le n$,
\eqn\sigdef{
\s_b(u_a) \equiv Q_{0,\ldots,\widehat{n-b},\ldots,n}(u_a),\qquad
0 \le b \le n,
}
are precisely the elementary symmetric polynomials in the $u_a$:
\eqn\sigmas{\eqalign{
\s_0 &= 1,\cr
\s_1 &= \sum_a u_a,\cr
\s_2 &= \sum_{a<b} u_a u_b,\cr
&\vdots\cr
\s_n &= \prod_{a=1}^n u_a.
}}
The $\s_a$ are good coordinates on 
$\Sym^k(\IC)$ in a neighborhood of $u_a=0$ and are therefore
the coordinates of interest as we take $n$ points $z_a$ to $z$
keeping the other $k-n$ points fixed. Note
that each $Q_{\{c_a\}}$ of degree greater than $n$
can be expressed as a polynomial in the $\s_a$.

The inverse of the Vandermonde matrix is
\eqn\vinv{
(V^{-1})^a{}_b = {(-1)^{a+b} \over \Delta} 
F_{0,\ldots,\widehat{b},\ldots,n}(u_1,\ldots,\widehat{u_a},\ldots,u_n).
}
Finally, we will use the identity
\eqn\identa{\eqalign{
Q_{0,\ldots,\hat{b},\ldots,n-1,p}(u_a) &=
{(-1)^{n} \over \Delta} 
\sum_{a=1}^n (-1)^au_a^p F_{0,\ldots,\widehat{b},\ldots,n-1}(u_1,\ldots,
\widehat{u_a},\ldots,u_n)\cr
&= (-1)^{n+b} \sum_{a=1}^n  u_a^p (V^{-1})^a{}_b
}}
which is obtained
by expanding out the determinant defined by the left-hand side
in the last row (the row containing $u_a^p$).

\subsec{Completeness}

We start with the basis $\{ |z_i \ra\}$.
Expanding out the exponential
in $|z_a\ra = e^{u_a \adag} |z\ra$
gives
\eqn\expa{
|z_a \ra = \sum_{b=1}^n |z;b\ra V^b{}_a + \sum_{p=n+1}^\infty
|z;p\ra u_a^{p-1}
}
where
\eqn\expb{
|z;m \ra \equiv {(\adag)^{m-1} \over (m-1)!}|z\ra.
}
Multiplying  \expa\ by $V^{-1}$ gives
\eqn\expc{\eqalign{
 |z_a \ra (V^{-1})^a{}_b &= |z;b\ra + \sum_{p=n+1}^\infty
|z;p\ra \sum_{a=1}^n u_a^{p-1} (V^{-1})^a{}_b
\cr
&= |z;b\ra + (-1)^{n+b} \sum_{p=n+1}^\infty
|z;p\ra Q_{0,\ldots,\widehat{b},\ldots,n-1,p-1}(u_a),
}}
using \identa.

The polynomial multiplying $|p\ra$ in \expc\ has degree greater than
zero for any $p>n$,
so that when we take all the $u_a$ to zero these terms
vanish and we have simply
\eqn\aaa{
\lim_{z_a \to z} |z_a \ra (V^{-1})^a{}_b =
|z;b \ra.
}
This establishes both \zlimit\ and the fact that the
Jacobian of the necessary change of basis is the
Vandermonde determinant, as advertised above \regkahler.

\subsec{Smoothness}

Let $y_i = z_i - c$ be the relative coordinates on $\CM_k$ as in section
3.3.  Repeated application of the exponential expansion yields the
formula
\eqn\kaha{
\det (e^{\bar{y}_i y_j}) = {1 \over k!} \sum_{n_1,\ldots,n_k=0}^\infty
{1 \over n_1! \cdots n_k!} \left| F_{n_1,\ldots,n_k}(y_1,\ldots,y_k)\right|^2.
}
Since each $F$ is divisible by the Vandermonde determinant
$\Delta = \prod_{i>j}(y_i-y_j)$ as in \qdef,
the K\"ahler potential \regkahler\ is
simply given by
\eqn\regkaha{\eqalign{
e^{K'} &= {1 \over k!}
\sum_{n_1,\ldots,n_k=0}^\infty
{1 \over n_1! \cdots n_k!} \left| Q_{n_1,\ldots,n_k}(y_1,\ldots,y_k)\right|^2
\cr
&= {1 \over k!} \left[
c_k(0) + \sum_{i=1}^k c_k(i) \left| \s_i(y) \right|^2 + \CO(\sigma_i^3)
\right],
}}
where the $c_k(i)$ are some positive numbers.
It is manifest from the second line of \regkaha\ that $K'$ is
well behaved as the solitons are brought together, i.e. as
$\s_i(y) \to 0$. We see directly from this derivation
how dividing $e^K$ by the Vandermonde determinant renders
the K\"ahler potential $K'$ nonsingular on the coincident locus.

\appendix{B}{Construction of the soliton on the integral torus}

We consider a torus with periodicities $l$ and $\tau l$ (in units of
$\sqrt{\t}$), where $A\equiv
\tau_2l^2/2\pi$ is assumed to be an integer. The 
generators of the fundamental group are represented by the unitary
operators 
\eqn\udefs{
U_1 \equiv e^{-il\hat y^2}, 
\qquad U_2 \equiv e^{il(\tau_2\hat y^1-\tau_1\hat y^2)},
}
which commute for such an integral torus. We wish to construct the
projection operator whose image is spanned by the lattice of coherent
states,
\eqn\latcoh{
U_1^{j_1}U_2^{j_2}|0\ra,
\qquad (j_1,j_2)\in\IZ^2.
}
Following the same strategy as on the cylinder, we will find a particular
linear combination,
\eqn\psilin{
|\psi\ra = \sum_{j_1,j_2}c_{j_1j_2}U_1^{j_1}U_2^{j_2}|0\ra,
}
that satisfies
\eqn\orthocon{
\la\psi|U_1^{j_1}U_2^{j_2}|\psi\ra = \delta_{j_10}\delta_{j_20}.
}
The projection operator we seek is then
\eqn\pfrompsi{
P = \sum_{j_1,j_2}U_1^{j_1}U_2^{j_2}|\psi\ra\la\psi|U_2^{-j_2}U_1^{-j_1}.
}

We employ a generalization of the so-called $kq$ representation
\refs{\bgz,\zak},
which provides a basis of simultaneous eigenstates of $U_1$ and $U_2$:
\eqn\kqdef{
|kq\ra \equiv 
\sqrt{{l\over2\pi}}e^{-i\tau_1(\hat y^2)^2/2\tau_2}\sum_je^{ijlk}|q+jl\ra,
}
where the ket on the right is a $\hat y^1$ eigenstate. We thus have
\eqn\uevalue{
U_1|kq\ra = e^{-ilk}|kq\ra, \qquad U_2|kq\ra = e^{il\tau_2q}|kq\ra.
}
The set $\{|kq\ra:0\le k<2\pi/l,0\le q<l\}$ forms an orthonormal and
complete basis for the Hilbert space. In terms of wave functions in the
$kq$ representation, \psilin\ becomes
\eqn\cdef{
C_\psi(k,q) \equiv \la kq|\psi\ra =
\sum_{j_1,j_2}c_{j_1j_2}e^{-ij_1lk+ij_2l\tau_2q}\la kq|0\ra
= \tilde c(k,q)C_0(k,q),
}
where
\eqn\howf{
C_0(k,q) \equiv \la kq|0\ra =
{1\over\pi^{1/4}\sqrt{l}}\exp(-{\tau\over2i\tau_2}k^2+ikq)
\vartheta_{00}({q+k\tau/\tau_2\over l},{\tau\over A}).
}
Note that $\tilde c$ is doubly periodic:
$\tilde c(k+2\pi/l,q) = \tilde c(k,q+l/A) = \tilde c(k,q)$.
The orthonormality condition \orthocon\ becomes
\eqn\orthokq{\eqalign{
\delta_{j_10}\delta_{j_20} =&
\int_0^{2\pi/l}dk\int_0^ldq\,
e^{-ij_1lk+ij_2l\tau_2q}\left|C_\psi(k,q)\right|^2 \cr
=& \int_0^{2\pi/l}dk\int_0^{l/A}dq\,
e^{-ij_1lk+ij_2l\tau_2q}\left|\tilde c(k,q)\right|^2
\sum_{n=0}^{A-1}\left|C_0(k,q+n{l\over A})\right|^2.
}}
Hence (choosing $\tilde c$ real),
\eqn\Cpsi{
C_\psi(k,q) =
{C_0(k,q) \over
\sqrt{2\pi/A\,\sum_{n=0}^{A-1}\left|C_0(k,q+nl/A)\right|^2}}.
}

With $|\psi\ra$ now in hand, we would like to find the field configuration
corresponding to the projection operator \pfrompsi. The inverse Weyl-Moyal
transformation yields a Fourier expansion on the torus:
\eqn\phifromp{\eqalign{
\ph(y_1,&y_2) = \cr
& {\l\over A}\sum_{j_1,j_2}
\la\psi|\exp\left[
{2\pi i\over l}\left(
j_1\hat y^1+{j_2-\tau_1j_1\over\tau_2}\hat y^2
\right)
\right]\!|\psi\ra
\exp\left[-{2\pi i\over l}\left(
j_1y^1+{j_2-\tau_1j_1\over \tau_2}y^2\right)\right].
}}
To find the Fourier coefficients in terms of the $kq$ wave function of
$|\psi\ra$, we need the following result, which may be derived from
\kqdef:
\eqn\kqfourier{\eqalign{
\la k'q'|\exp&\left[
{2\pi i\over l}\left(
j_1\hat y^1+{j_2-\tau_1j_1\over\tau_2}\hat y^2
\right)
\right]\!|kq\ra
= \cr
& \sum_j
\exp\left({2\pi ij_1q'\over l}+{\pi ij_1j_2\over A}+ijlk\right)
\delta(q-q'-{j_2l\over A}+jl)
\sum_{j'}\delta(k-k'-{2\pi j'\over l}).
}}
One can then show that
\eqn\phifromC{\eqalign{
\ph(y^1,y^2)&= 
{\pi\l\over A}\sum_{j_2=0}^{2A-1}
e^{-2\pi ij_2y^2/(\tau_2l)}
\left(
C_\psi^*(y^2,y^1-{\tau_1\over\tau_2}y^2-{j_2l\over2A})
C_\psi(y^2,y^1-{\tau_1\over\tau_2}y^2+{j_2l\over2A})
\right.\cr {} & \quad\qquad + \left.
C_\psi^*(y^2+{\pi\over l},y^1-{\tau_1\over\tau_2}y^2-{j_2l\over2A})
C_\psi(y^2+{\pi\over l},y^1-{\tau_1\over\tau_2}y^2+{j_2l\over2A})
\right).
}}
Combining this formula with \Cpsi\ and \howf, one finally arrives at the
expression \torus.

\listrefs

\end